\documentstyle[11pt,paspconf,epsf]{article}
\def\cm3{cm$^{-3}$}

\begin{document}

\title{Progress in Model Atmosphere Studies of WR stars}

\author{Paul A. Crowther}
\affil{Dept of Physics \& Astronomy, UCL, Gower St. London~WC1E~6BT}

\begin{abstract}
Recent progress in the quantitative analysis of Wolf-Rayet stars
is reviewed, emphasising the role played by choice of spectral 
diagnostics, clumping and line blanketing on derived stellar properties.
The ionizing properties of WR stars are discussed, based on clumped, 
line blanketed models for WN and WC stars. The role of metallicity 
and mass-loss is assessed, and the role of H\,{\sc ii} regions 
as probes of predicted Lyman continuum distributions. Suggestions are 
made for differences in observed properties of WCE and WO subtypes.
\end{abstract}


\section{Introduction}

It is only through understanding the physics
of massive stars, their atmospheres, radiation, and evolution,
that we will be able to make progress in many aspects of 
astrophysics. Particularly important is the quantitative study of young 
starbursts, which are dominated by the effects of O-type and Wolf-Rayet (WR) 
stars. WR stars comprise only 10\% of the massive stellar content 
in the Galactic mini-starburst, NGC\,3603, yet they contribute
20\% of the total ionizing flux and 60\% of the total kinetic
energy injected into the ISM (Crowther \& Dessart 1998).
In order that young starbursts can be properly studied, both nearby, 
and at high-redshift, it is crucial that the properties and 
evolution of O-type and WR stars spanning a range of initial 
metallicities are determined. In this review, I will consider 
recent theoretical progress in WR analyses that has been made 
towards this goal, focusing especially on spectroscopic and 
ionizing properties.

\section{Quantitative spectroscopy of WR stars}

Quantitative analysis of W-R stars represents a formidable challenge, 
since their stellar winds are so dense that their 
photospheres are invisible, so that the usual assumptions of plane-parallel 
geometry and local thermodynamic equilibrium (LTE) are wholly inadequate.
A minimum requirement is to consider non-LTE in an extended, expanding 
atmosphere for multi-level atoms. At present, three independent model 
atmosphere codes are capable of routinely analysing the spectra of 
WR stars, considering complex model atoms of hydrogen, helium, nitrogen,
carbon and oxygen, developed by W.-R.~Hamann (Potsdam), W.~Schmutz (Zurich)
and D.J.~Hillier (Pittsburgh), the latter also implemented by P.A.~Crowther 
(London) and F.~Najarro (Madrid). Each code solves the radiative transfer 
problem in the co-moving frame, subject to statistical and radiative 
equilibrium, including the effects of electron scattering and clumping. 
Overall, consistency between results from these codes is very good.
Schmutz and Hillier have also accounted for line blanketing by heavy 
elements.

Individual calculations are computationally demanding, so that a large 
parameter space can not be readily explored. Consequently, 
computationally quick codes have been developed which solve the 
transfer problem in the Sobolev approximation and assume a
wind temperature distribution (e.g. de Koter et al. 1997; Machado, 
these proc.)  De Marco et al. (these proc.) compare the predictions of 
the code by de Koter with that of Hamann for WC stars.

\section{Progress in the determination of stellar properties}

In the simplest case, the stellar properties of WR stars are derived 
from the following diagnostics: two spectral lines from adjacent ionization 
stages of helium (He\,{\sc ii} $\lambda$5412 and He\,{\sc i} $\lambda$5876 most 
commonly), plus the absolute magnitude in a standard filter (typically 
$M_{v}$) and the terminal wind velocity ($v_{\infty}$), often measured 
from UV resonance lines. The default number of model parameters available is 
therefore four, $R_{\ast}$, $T_{\ast}$, $\dot{M}$ and $v_{\infty}$. 
Stellar temperatures for extended atmospheres are related to the 
inner boundary of the model atmosphere (generally around Rosseland 
optical depth $\tau_{\rm Ross}$$\sim$20), 
which  often deviates significantly from the `effective' temperature, 
at $\tau_{\rm Ross}$=2/3 (Hamann 1994). Schmutz et al. (1989) identified the
so-called transformed radius ($R_{t}$), a measure of wind density,  which
relates $R_{\ast}$, $\dot{M}$ and $v_{\infty}$ so that almost identical
spectra are produced for fixed $R_{t}$, reducing the number of free 
parameters. In this way, a large number of WR stars in the Galaxy and 
Magellanic Clouds have been analysed by Hamann and co-workers,
comparing observed line equivalent widths to interpolations of large model 
grids (see Hamann, these proc.). However, actual WR stars are not 
pure helium stars, so that the contribution of other elements is necessary. 

\subsection{The effect of including metals}

It was soon established that the wind properties of WR stars were affected
by the presence of metals, notably carbon and nitrogen (Hillier 1988; 1989).
For WN stars, metals are trace elements ($\sim$1\% in nitrogen by mass),
so pure He analyses compare relatively well with those additionally including
carbon and nitrogen, which control the outer wind properties. In late-type
WN (WNL) stars hydrogen contributes significantly (up to $\sim$50\% in 
extreme cases). Determination of atmospheric contents requires detailed 
analysis of individual stars through a comparison between theoretical line 
profiles (e.g. H$\alpha$ for hydrogen) and spectroscopic observations 
(e.g. Crowther et al. 1995a). In WC stars, it was soon realised
that pure He studies were inadequate to obtain reliable stellar properties,
since carbon mass fractions are $\sim$40--50\% (Hillier 1989). WC analyses
need to use He and C diagnostics simultaneously in order that the
stellar and chemical properties may be determined. The degree of complexity 
in atomic data handled for carbon has a great influence on predicted 
line strengths (Hillier  \& Miller 1999).

\begin{figure}[thbp]
\epsfxsize=13.0cm \epsfbox[30 285 530 510]{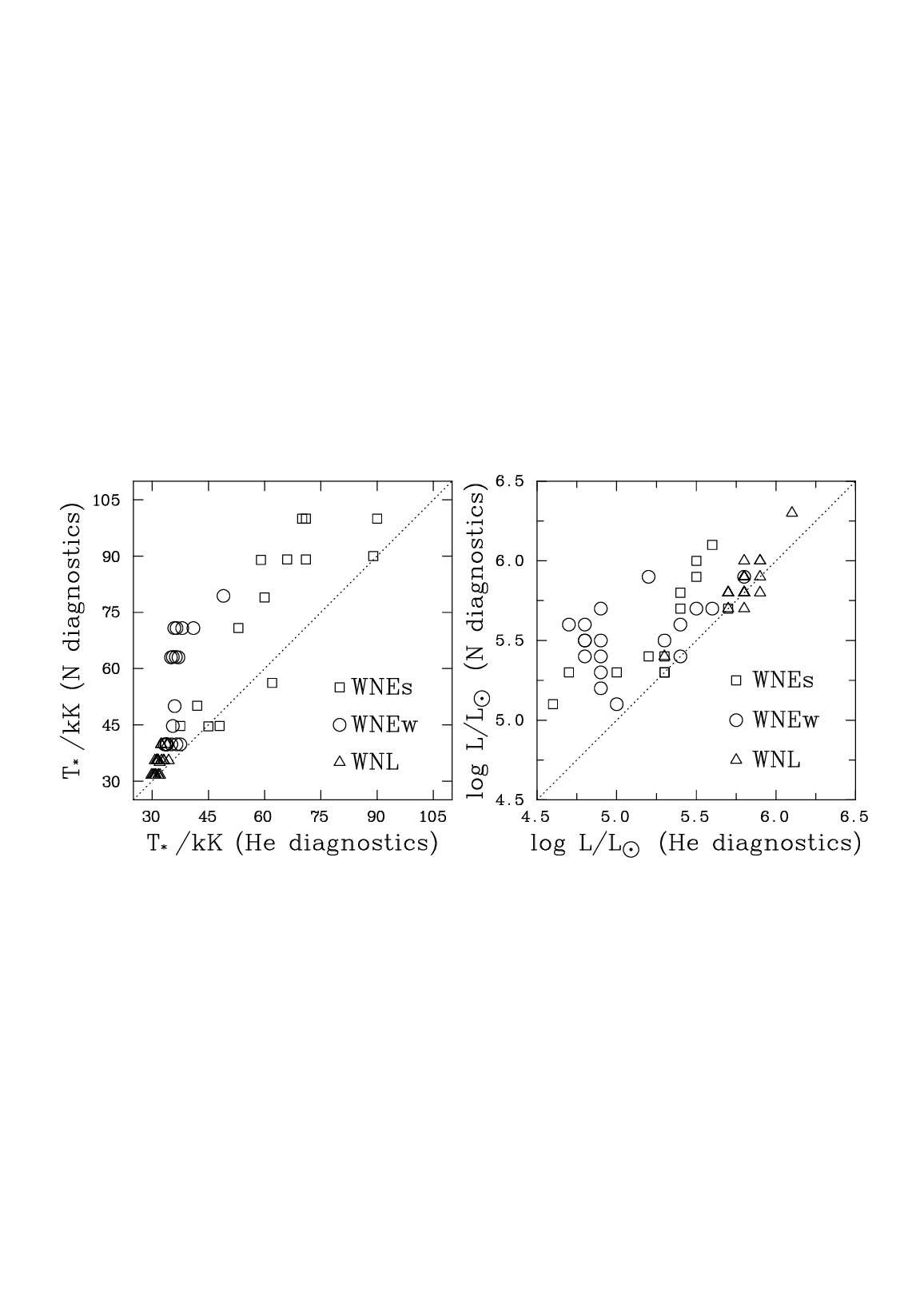}
\caption{Comparison between stellar temperatures and luminosities
of Galactic WN stars  obtained from optical helium diagnostics 
(Hamann et al. 1995) and nitrogen diagnostics (Hamann \& Koesterke 1998a),
adjusted for the assumed absolute magnitudes of the latter work.
Larger stellar temperatures imply dramatically higher luminosities 
for weak lined WNE stars.}
\label{fig1}
\end{figure}

This technique has a major disadvantage in that it is time consuming and
computationally expensive. Nevertheless, derived stellar properties can be
considered robust, while model deficiencies may be readily identified. 
Studies of individual stars have now been performed for Galactic, Magellanic Cloud
and other Local Group WR stars using 4m class ground-based telescopes 
(e.g. Smith et al. 1995 in M33). Source confusion becomes problematic at 
large distances from ground-based observations (0.$''$8 corresponds to a 
scale of 3 parsec at the modest distance of M33).

\subsection{Choice of spectral diagnostics}

Although helium and carbon diagnostics are combined to derive the
properties of WC stars, the majority of WN studies use solely helium.
Early results for weak-lined, early-type WN (WNE) stars led to surprisingly
low stellar temperatures (e.g. Hamann et al. 1995), which were comparable 
with WNL stars instead of strong-lined WNE stars. Crowther et al. (1995b)
demonstrated that the stellar temperatures of weak-lined WNE stars are
in line with strong-lined examples, if nitrogen diagnostics 
(e.g. N\,{\sc iv} $\lambda$4058, N\,{\sc v}  $\lambda$4603--20) are used 
instead of helium. Results from helium are more straightforward, since the
availability and quality of its atomic data is superior to nitrogen. 
However, He\,{\sc i} lines are typically formed at large radii from the core, 
and are extremely weak in hot WR stars, with the exception of 10830\AA\ that is 
observationally challenging. In contrast, 
N\,{\sc iv-v} lines originate from the inner 
wind and are readily observed. Consequently, nitrogen ought to serve as a 
more sensitive diagnostic of the stellar temperature, and circumvent the
problem identified by Moffat \& Marchenko (1996). They noted that 
stellar radii derived from He analyses were {\it greater} than the 
orbital radii of some short period WR+O binaries. Discrepancies for 
WN stars are not restricted to He and nitrogen diagnostics (see e.g. 
Crowther \& Dessart 1998).

Hamann \& Koesterke (1998a) have recently re-analysed a large sample 
of Galactic WN stars and arrived at similar conclusions to Crowther et
al. (1995b). In Fig.~\ref{fig1}, the stellar temperatures and 
luminosities of WN stars obtained by alternative helium and nitrogen 
diagnostics are compared. 
Consistent results are obtained for WNL stars, while higher 
temperatures  and luminosities are obtained from nitrogen 
diagnostics for WNE stars, especially weak-lined stars. Differences in 
derived temperatures may be large, increasing by a factor of up to
two (from 36kK to 71kK for the WN5(h) star WR49). The change in 
luminosity is greater still -- increasing by a factor of six in this 
star because of the sensitive dependence of bolometric correction (B.C.) with 
temperature. Parameters derived from helium or nitrogen lines should be fully 
consistent, so that discrepancies indicate that something is missing in
current models. Perhaps clumps in the wind affect the ionization balance in 
the He\,{\sc i} line forming region of WNE stars -- this may also be
relevant to the (poorly predicted) strength of He\,{\sc i} P~Cygni absorption 
components. {\it Whatever the cause, care should be taken when comparing 
results obtained from different spectral diagnostics}.
 
Detailed analyses of WC-type stars have also been carried out (e.g.
Koesterke \& Hamann 1995; Gr\"{a}fener et al. 1998) using helium, carbon 
and occasionally oxygen diagnostics. However, the additional number of free 
parameters (C/He, O/He) has restricted the sample analysed to date, and
oxygen diagnostics lie in the near-UV, requiring space based observations
 (Hillier \& Miller 1999). Since the UV and optical spectra of WC stars are 
dominated by overlapping broad emission lines, it is difficult to assign 
suitable continuum windows. Analyses typically consider the continuum 
and line spectra in isolation, namely that interstellar extinctions are obtained 
by  matching continua to de-reddened observations, while theoretical line 
profiles are compared to normalized spectra. A less error-prone approach is the 
comparison between de-reddened fluxed observations and synthetic spectra, accounting 
for line overlap. In this way, erroneously defined continuum windows (e.g.
at He\,{\sc ii} $\lambda$5412, Hillier \& Miller 1999), and 
unusual UV extinction laws, may be identified. 

\subsection{IR analyses}

Studies discussed above rely exclusively on optical (or  occasionally
UV) spectral diagnostics. The first infra-red (IR) spectroscopy of WR stars was 
obtained by Williams (1982), although recent advances mean that this 
wavelength region can now be used to observe a large sample of WR stars, 
particularly those obscured at shorter wavelengths. 
Crowther \& Smith (1996) have assessed the reliability of IR analyses 
of WR stars by studying two WNE stars for which UV and optical data sets 
were also available. They found that results from exclusively near-IR 
observations were in good agreement with optical studies, and with later
Infra-red Space Observatory (ISO) spectroscopy for WR136 (R.~Ignace, priv. 
comm.). Bohannan \& Crowther (1999) have recently compared optical and IR
analyses of Of and WNL stars.

Quantitative IR studies of WN-like stars at our Galactic Centre  have recently
been presented (e.g. Najarro et al. 1997a). 
Unfortunately, the majority of these stars are relatively cool, so that
the sole K-band He\,{\sc ii} diagnostic at 2.189$\mu$m is unavailable.
Without a second ionization stage, a unique temperature may not be obtained,
so that mass-loss rates and abundances are uncertain. Dessart et al. 
(these proc.) attempt to solve this by using the stronger He\,{\sc ii} 
3.09$\mu$m line in the thermal IR as a temperature diagnostic. Another
limitation with the K-band is that the prominent He\,{\sc i} line at 
$\lambda$2.058$\mu$m is strongly affected by (metallicity dependent) line 
blanketing effects, as shown by Crowther et al. (1995a, 1998). 
Problems with the quantitative analysis of low temperature stars are neatly 
summarised by Hillier et al. (1999) for the Galactic early B-type supergiant 
HDE\,316285.
They obtained a wide range of possible mass-loss rates and surface H/He 
abundances for this star, despite the availability of high quality optical and near-IR 
spectroscopy.

\section{Relaxing the standard assumptions}

Model calculations so far discussed use
$R_{\ast}$, $T_{\ast}$, $\dot{M}$ and $v_{\infty}$,
plus elemental abundances as free parameters. However, observational
evidence suggests that presently assumed quantities, such as the
velocity law and homogeneity may be inappropriate. In addition,
it is well known that line blanketing by thousands of transitions in 
the ultraviolet (UV) and extreme ultraviolet (EUV) need to be incorporated
into calculations. 
Each additional relaxation adds (at least) one new 
variable parameter to the existing set. Consequently,
of the several hundred WR stars thus far analysed quantitatively, to date 
studies of only two have included assorted elements, a variety of 
velocity laws, clumping and line blanketing (Schmutz 1997; Hillier \& 
Miller 1999).

\subsection{Variations in velocity law}

Generally, a uniform form of the radial velocity field 
($v \propto v_{\infty}(1-R/r)^\beta$) is assumed, of exponent 
$\beta$=1. Tailored analysis are 
required to test alternative laws. Unfortunately, different velocity 
forms are frequently able to reproduce the observed spectrum equally 
well (Hillier 1991a). 
In some cases, specific exponents produce optimum agreement, provided with
a  suitably large range of spectroscopic observations. From a careful 
comparison of the optical and far-red appearance of the Luminous Blue 
Variable (LBV) P~Cygni, Najarro et al. (1997b) found that a $\beta$=4.5 law 
provided the best match. Including mid-IR ISO 
observations led to a revision to $\beta$=2.5 (Najarro et al. 1998). 
Unfortunately, a long wavelength observational baseline is rarely available. 
Schmutz (1997) went a stage further by {\it deriving} the 
form of the velocity law in WR6 from hydrodynamics, at least in
the outer visible part of the wind, with $\beta$=3. As a indication of the
reliability of this approach, the emission profile of 
He\,{\sc i} $\lambda$10830 was reproduced better than in previous studies.

\subsection{Wind inhomogeneities and departures from spherical symmetry}

WR winds are known to be inhomogeneous, from both observational and
theoretical arguments (Willis, these proc.). However, homogeneity
has been assumed by the majority of atmosphere studies to date. 
Consideration of electron scattering -- causing a frequency redistribution of 
line photons -- provides the key to spectral synthesis (Hillier 1984; 1991b).
Homogeneous models often overestimate the strength of electron scattering 
wings relative to the overall emission line intensity. Since free-free 
emission and radiative recombination both scale as the square of 
the density, whereas the electron scattering opacity scales linearly with 
density, estimates of wind inhomogenities may be estimated by varying 
volume filling factors and mass-loss rates so that line profiles and 
electron scattering wings are simultaneously reproduced. In line transfer 
calculations performed to date, several simplifying assumptions are 
made, namely that models are composed of radial shells of material, with
no inter-clump medium. The variation of clumping factor with radius 
taken into consideration in some cases since radiative instabilities
are not expected to be important in the inner wind. 

Schmutz (1997) and 
Hillier \& Miller (1999) have estimated mass-loss rates of 
WR6 and WR111 which are a factor of 3--4 lower than 
those resulting from homogeneous models. Hamann \& Koesterke (1998b) have also 
applied an identical approach to a sample of four WR stars, with similar 
results obtained. Substantially lower mass-loss rates of WR stars 
has importance in evolutionary model calculations and in reducing the
momentum (alternatively opacity) problem of driving 
WR winds (Gayley et al. 1995).

\begin{figure}[thbp]
\epsfxsize=13.0cm \epsfbox[15 285 515 510]{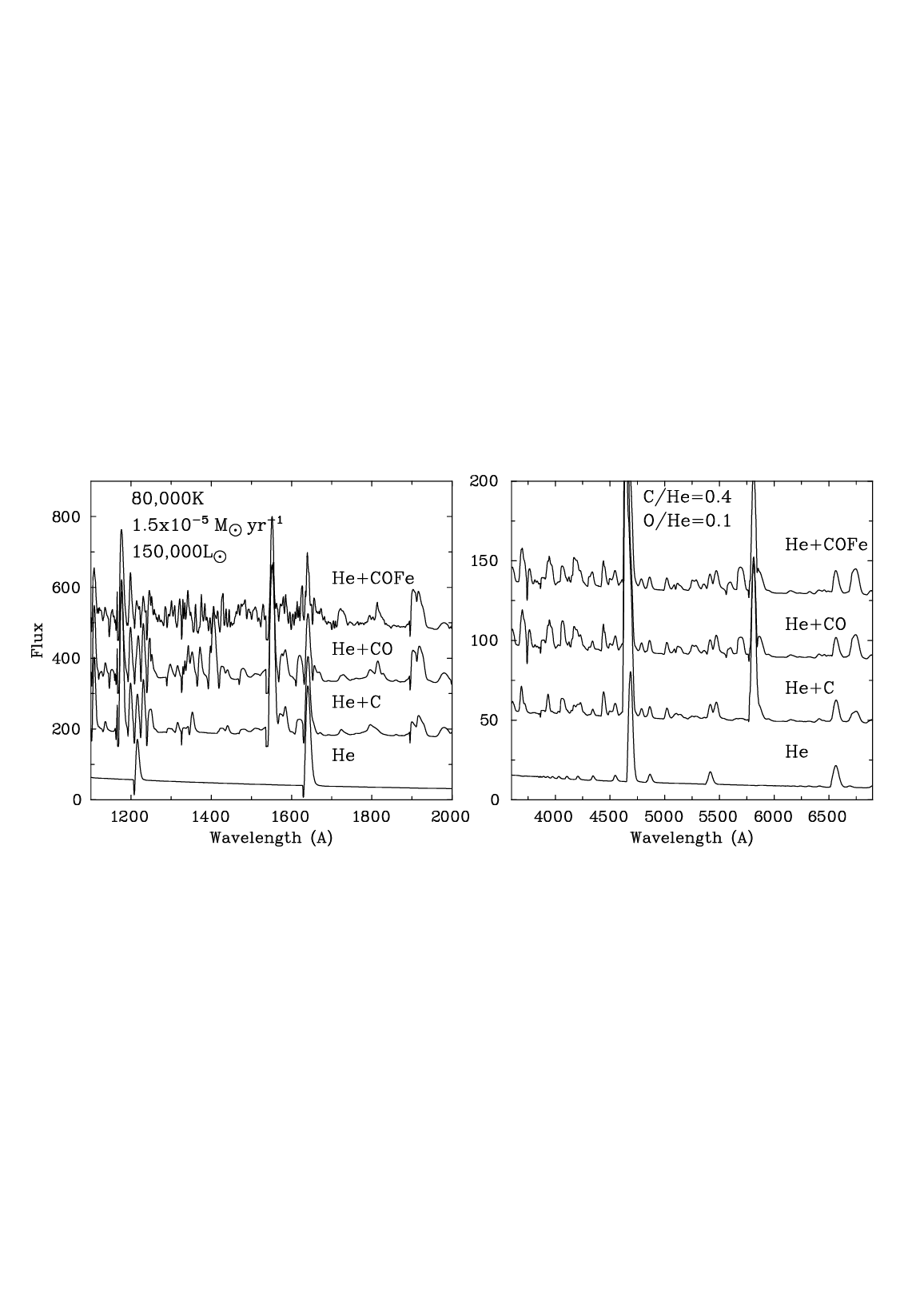}
\caption{Comparison between a WC model at fixed stellar parameters,
including He, He+C, He+C+O and finally He, C, O and Fe using the
Hillier \& Miller (1998) code.}
\label{fig2}
\end{figure}

To date, all spectroscopic studies have assumed spherical symmetry. 
Evidence from spectropolarimetry indicates that this is appropriate for 
$\sim$85\% of cases (Harries et al. 1998). For the remaining stars, density 
ratios of  2--3 are implied from observations. Future calculations 
will need to consider departures from spherical symmetry. Indeed,
the wind of the prototypical WNE star WR6 is grossly asymmetric.

\subsection{Influence of line blanketing}\label{4.3}

Observations of the forest of iron lines in the UV spectra of WR stars,
demonstrate the large influence that line blanketing by Fe-group elements 
has on the emergent spectrum. The neglect of blanketing reveals itself 
through inconsistencies of model fits, and results from comparison with 
ionized nebulae. The principal problem in accounting for 
line blanketing is being able to 
handle the effect of tens of thousands of line transitions in the 
radiative transfer calculations. To date, solely Schmutz and Hillier have
made allowance for blanketing, albeit using different techniques, 
each with their own advantages and disadvantages. Monte Carlo sampling 
by Schmutz (1997) allows the opacity of a huge number of lines to 
be considered,  although spectral synthesis of individual features in 
the UV is not possible, while the reverse is true for Hillier \& Miller 
(1998) who use a `super-level' approach, treating the transfer problem
correctly. 

In Fig.~\ref{fig2} models for a WCE star are compared, in which 
increasing number 
of elements are included, He, C, O, and Fe. Carbon and oxygen 
have a considerable effect on the UV and optical energy distribution of the 
models, with Fe modifying the energent UV flux distribution
(Hillier \& Miller 1998, 1999). What effect does allowing for 
clumping and line blanketing have on the
resulting stellar properties? In Table~\ref{table1} the
results of Schmutz (1997) and Hillier \& Miller (1999) for WR6 (WN4b)
and WR111 (WC5) are compared with earlier studies. 
Stellar temperatures and bolometric
luminosities of the blanketed analyses are considerably greater 
than those from unblanketed models, with a significant EUV excess (and
corresponding increase of B.C.), while mass-loss rates are significantly
lower, as a result of considering clumped winds. For the case 
of WNL stars, Crowther et al. (1998) and Herald et al. (these proc.)
find that blanketing has a minor influence on stellar temperatures
(though the EUV energy distribution is affected). This result is in
apparent contradiction with the analysis of LMC WN9--11 stars by Pasquali et al. 
(1997) using {\it grids} of line blanketed models. Pasquali et al.
revealed considerably higher temperatures relative to earlier unblanketed 
tailored analyses (Crowther et al. 1995a; Crowther \& Smith 1997). Subsequent 
{\it tailored} spectroscopic analyses including blanketing by Pasquali 
(priv. comm.), agree well with the parameters obtained by Crowther \& Smith.

 \begin{table}
 \caption{Comparison of WR stellar properties derived from recent 
spectroscopic analyses including
blanketing and clumping (Schmutz 1997 S97; Hillier \& Miller 1999, HM99) 
relative to earlier studies not accounting for these effects
(Schmutz et al. 1989 SHW89; Koesterke \& Hamann 1995 KH95; Hamann et al. 1995
HKW95)}
\label{table1}
 \begin{center}\scriptsize
 \begin{tabular}{l@{\hspace{7mm}}l@{\hspace{7mm}}l@{\hspace{7mm}}
l@{\hspace{7mm}}l@{\hspace{7mm}}l@{\hspace{7mm}}l}
$T_{\ast}$ & log~$L$ & log~$\dot{M}$ & B.C. & $\underline{\dot{M}v_{\infty}}$
&Diagnostics&Ref.\\
kK       & $L_{\odot}$ & $M_{\odot}$yr$^{-1}$ & mag& $L/c$ &\\
\noalign{\smallskip}
 \tableline
\noalign{\smallskip}
   \multicolumn{5}{c}{WR6 = HD50896 (WN5 or WN4b)} \\
71      & 5.2 & $-$4.1 &$-$3.7 & 37 &He  & HKW95 \\
84      & 5.7 & $-$4.5 (cl)&$-$4.9 &  6 &He  & S97 \\
\noalign{\smallskip}
   \multicolumn{5}{c}{WR111 = HD165763 (WC5)} \\
35      & 4.6 & $-$4.6 &$-$3.2& 90 &He & SHW89 \\
62      & 5.0 & $-$4.3 &$-$4.0& 50 &He+C & KH95 \\
90      & 5.3 & $-$4.8 (cl) &$-$4.7 & 10 &He+C+O & HM99\\
\noalign{\smallskip}
%\tableline
 \end{tabular}
 \end{center}
 \end{table}

Our ability to synthesise individual and groups of Fe lines in the spectrum
of WR stars suggests that they can be used to derive Fe-group abundances. 
UV spectra of O stars (Haser et al. 1998) and optical spectra of A~supergiants 
(McCarthy et al. 1995) have previously been used to determine
Fe-contents in extra-galactic environments, though few detailed attempts have
been made using WR stars (see Hillier \& Miller 1999). As an indication of 
the potential for the future, Crowther et al. (1999) have recently used 
Hubble Space Telescope (HST) spectroscopy of the erupting 
LBV V1 in the giant H\,{\sc ii} region NGC\,2363, within the 
Magellanic irregular galaxy NGC\,2366 (3.5\,Mpc) to 
determine its Fe-abundance.

\section{What are the ionizing spectra of Wolf-Rayet stars?}

The ionizing flux distribution of WR stars has importance in
the study of extra-galactic regions containing young massive stars 
(giant H\,{\sc ii} regions, WR galaxies etc.). Recent results for O 
stars incorporating non-LTE and wind effects have resulted in 
improved agreement with observations of associated H\,{\sc ii} regions 
(e.g., Stasi\'{n}ska \& Schaerer 1997). It is equally important that
suitable ionizing distributions for WR stars are used, which affect
determinations of IMFs and ages. Since the Lyman continuum distributions 
of WR stars is not directly visible (due to absorption by intervening 
hydrogen), indirect methods need to be used to verify current models.

\begin{figure}[thbp]
\epsfysize=17.0cm \epsfbox[-20 40 400 780]{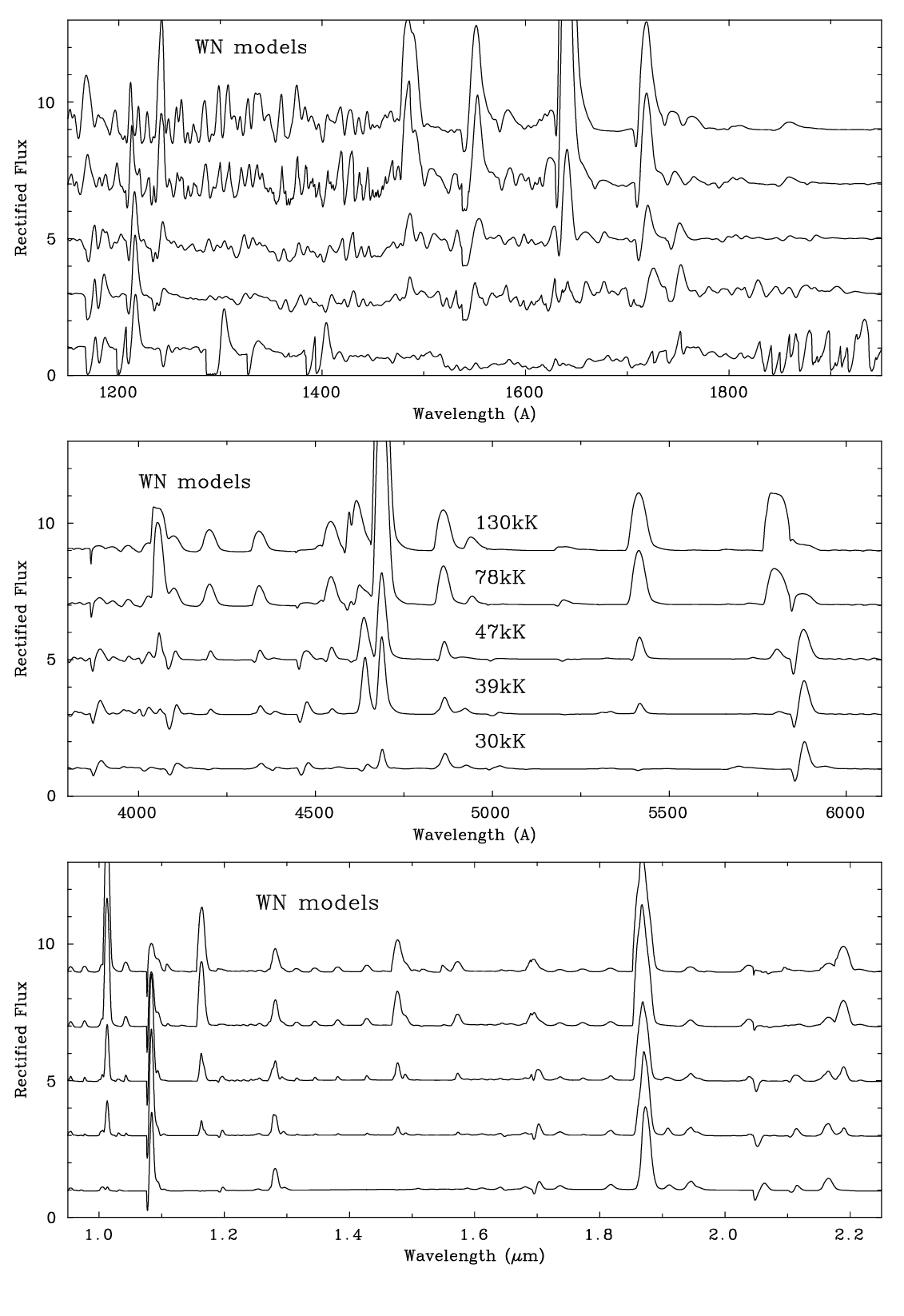}
\caption{Synthetic UV, optical and near-IR spectra of WN stars for 
a variety of
temperatures using the Hillier \& Miller (1998) line blanketing code,
spanning WN4 to WN9. Assumed parameters are 
log~($\dot{M}/M_{\odot}$yr$^{-1}$)=$-$4.6, 
log~$L_{\ast}/L_{\odot}$=5.5, $v_{\infty}$=2000\,km\,s$^{-1}$, 
plus H:He:C:N:Si:Fe mass-fractions (in \%) of 
0.2: 98: 0.03: 1.4: 0.09: 0.15 and a filling factor of 0.1.}
\label{fig3}
\end{figure}

\subsection{Nebulae as probes of the Lyman continuum flux distribution}

The principal work in this field was that of Esteban et~al. (1993) who
combined pure helium, unblanketed WR model fluxes (Schmutz et al. 1992) 
with observed properties of WR ring nebulae, to investigate the properties 
of the central stars. Esteban et al. varied stellar temperatures until agreement 
was reached between the observed and predicted nebular properties.
Comparisons with (independent) stellar analyses of the central stars
was found to be reasonable, except that lower temperatures were required from 
the photo-ionization models for WNL stars. Recently, Crowther et al. (1998) 
and Pasquali et al. (these proc.) have returned to 
this technique, newly considering the influence of line blanketing 
using the Hillier and Schmutz codes. They depart from Esteban et al. 
in  that ionizing flux distributions {\it obtained} from a stellar 
analysis of the central star are used in the photo-ionization modelling.

Crowther et al. (1998) compared line blanketed and unblanketed 
flux distributions resulting from stellar analyses of the WN8 star 
WR124, with observations of its associated nebula, M1--67. 
They found that the blanketed model predicted the nebular temperature 
and ionization balance much better than the unblanketed case. Allowance 
for improved nebular properties of M1--67 from Grosdidier et al. (1998), 
particularly the  radial density distribution, leads to even better 
agreement with observations. Pasquali et al. (these proc.) find good
agreement between the predicted and observed nebular properties of 
NGC\,3199, using stellar flux distributions from analyses of its
central WNE star WR18 with the Schmutz and Hillier codes. Unfortunately, the
observed properties ($T_{e}$, $N_{e}$, $\Delta R$, abundances etc.) of most 
WR nebulae at present are insufficiently well determined to use as tests
of stellar models. 

\begin{figure}[thbp]
\epsfxsize=13.0cm \epsfbox[30 285 530 510]{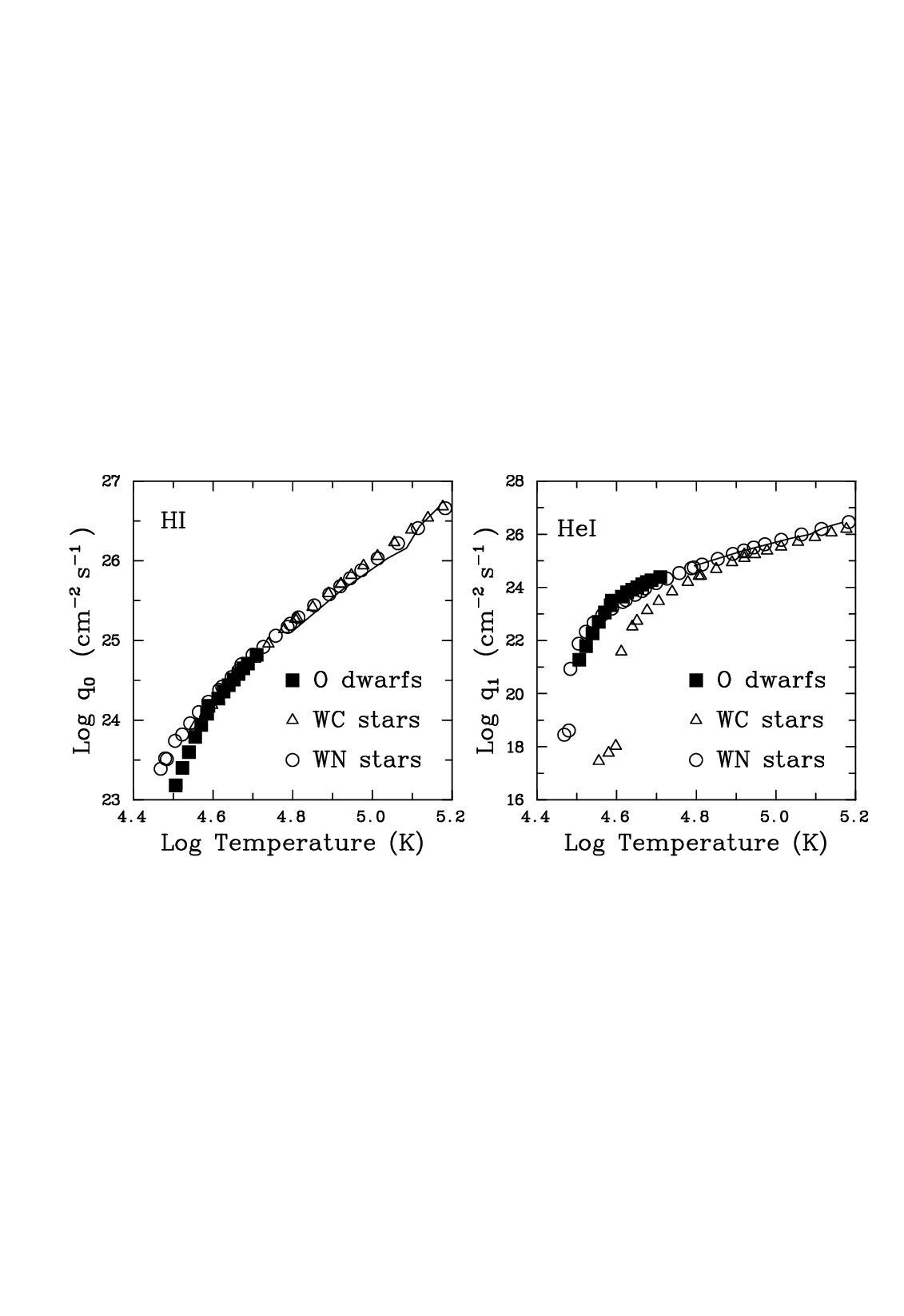}
\caption{Comparison between ionizing fluxes (photon\,s$^{-1}$\,cm$^{-2}$)
of line blanketed WN and WC models with O stars 
(Schaerer \& de Koter 1997),
and pure helium WR models of Schmutz et al. (1992, solid line).}
\label{fig4}
\end{figure}

\subsection{The effect of blanketing on ionizing fluxes}

Overall, spectral synthesis and photo-ionization modelling results give 
us confidence in the validity of current line  blanketed Wolf-Rayet codes. 
Since the only generally available WR models are unblanketed, pure helium energy 
distributions of Schmutz et al. (1992), how do new results compare? The calculation 
of a large multi-parameter grid of line blanketed models is a formidable 
computational challenge. For the moment, I have obtained models for WR stars with
the Hillier \& Miller (1998) code, varying solely temperatures (30 to 150kK).
WN models span WN4 to WN9 spectral types and include the 
effects of complex model atoms of H\,{\sc i}, He\,{\sc i-ii}, C\,{\sc ii-iv}, N\,{\sc ii-v},
Si\,{\sc iii-iv} and Fe\,{\sc iii-vii}. In Fig.~\ref{fig3}, selected 
synthetic UV, optical and near-IR spectra are presented. Similar calculations
for WC stars spanned WC4 to WC9 and included He\,{\sc i-ii},
C\,{\sc ii-iv}, O\,{\sc ii-vi} and Fe\,{\sc iii-vii} in detail.
Their predicted Lyman continuum distributions are 
fairly soft in all cases, with negligible emission above the He$^{+}$ edge at 54eV.
 
In Fig.~\ref{fig4} the ionizing fluxes of these models in
the H$^{0}$ and He$^{0}$ continua (in units of photon\,s$^{-1}$\,cm$^{-2}$)
are compared with 
recent solar metallicity O-star models (Schaerer \& de Koter 1997), 
plus the pure helium Schmutz et al. (1992) models. The 
line blanketed WN flux distributions support the pure helium Schmutz 
et al. (1992) predictions, although the additional blanketing from 
C and O in WC stars produces a softer ionizing spectrum at an identical 
temperature, with negligible flux emitted $\lambda \le$300\AA. WR 
stars also compare closely with comparable temperature O stars in 
their ionizing flux per unit area. 

\begin{figure}[thbp]
\epsfxsize=13.0cm \epsfbox[15 285 515 510]{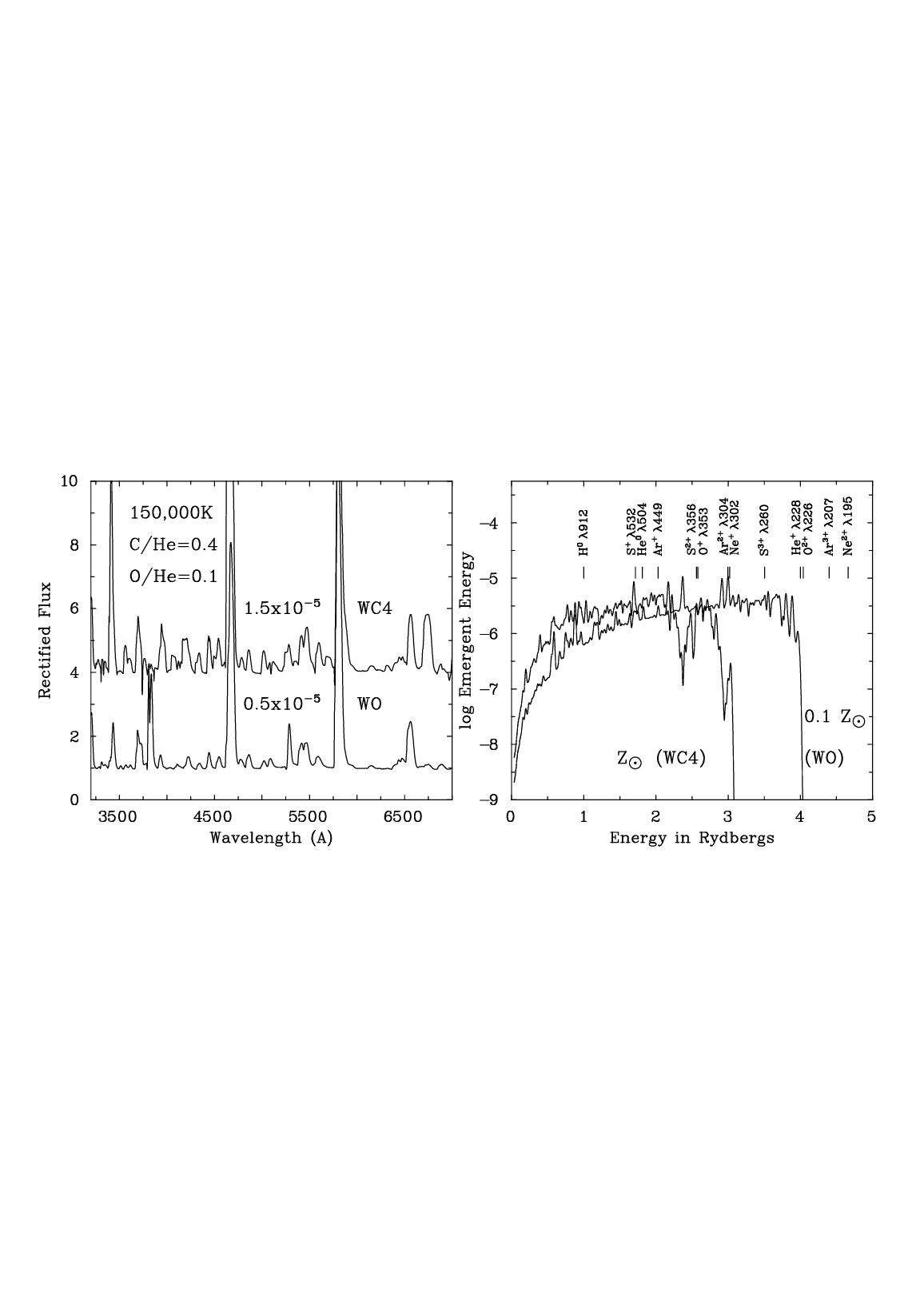}
\caption{Comparison between WCE models at fixed stellar parameters
(150kK, C/He=0.4, O/He=0.1), except that mass-loss rates (and Fe-contents)
differ by a factor of three (ten). The low metallicity/mass-loss model
would be classified as a WO-type star instead of a WC4 star, despite its 
identical atmospheric composition.}
\label{fig5}
\end{figure}

\subsection{The effect of wind density}

Schmutz et al. (1992) stress the importance of stellar wind density on
the ionizing flux distributions of WR stars, such that emission at 
$\lambda \le$228\AA\ relies on the WR wind being relatively transparent. 
Denser winds, such as those presented above for representative
Galactic WR stars, destroy photons beyond this edge. To illustrate this,
additional calculations have been performed for lower wind densities. 
Although a mass-loss versus metallicity ($Z$) scaling for WR stars has 
not been identified, let us assume that their winds are radiatively driven 
with a dependance of $\dot{M} \propto Z^{0.5}$ (as obtained for 
radiatively driven O-type stars by Kudritzki et al. 1989). 

I have taken the 150kK WC model, whose synthetic spectrum
approximates a WC4-type star, and solely reduced its mass-loss rate (by a factor 
of three) and Fe-content (by a factor of ten). The optical and ionizing spectrum
of the low wind density model are compared with the WC4 model in Fig.~\ref{fig5}, 
revealing a harder flux distribution (increasing the B.C. by 1.2 mag), and 
a dramatic change in the emergent optical spectrum. O\,{\sc vi} emission
is very strong so the low wind density case resembles a WO-type star.
Consequently, {\it a modest change in mass-loss rate has a major influence 
on the ionizing energy distribution and observed spectral appearance.}
Strong  O\,{\sc vi} emission in a WR spectrum is connected principally with the 
wind density, rather than elemental abundance (Smith \& Maeder 1991 identified 
WO stars as the chemically evolved descendants of WC stars). In WC4 stars, 
the high wind density and consequently very efficient wind cooling means 
that O$^{6+}$ recombines to O$^{5+}$  and subsequently O$^{4+}$ {\it interior} 
to the optical line formation region, producing observed O\,{\sc iv-v} lines. The 
less efficient cooling of WO winds, through a lower wind density (because of 
lower mass-loss rates and higher wind velocities) permits O$^{6+}$ 
recombination in the optical line  formation region, producing strong O\,{\sc vi} emission.
In support of this, recall that WO stars outnumber the WC population 
at low metallicities (SMC, IC1613).

\begin{figure}[thbp]
\epsfxsize=13.0cm \epsfbox[15 285 515 510]{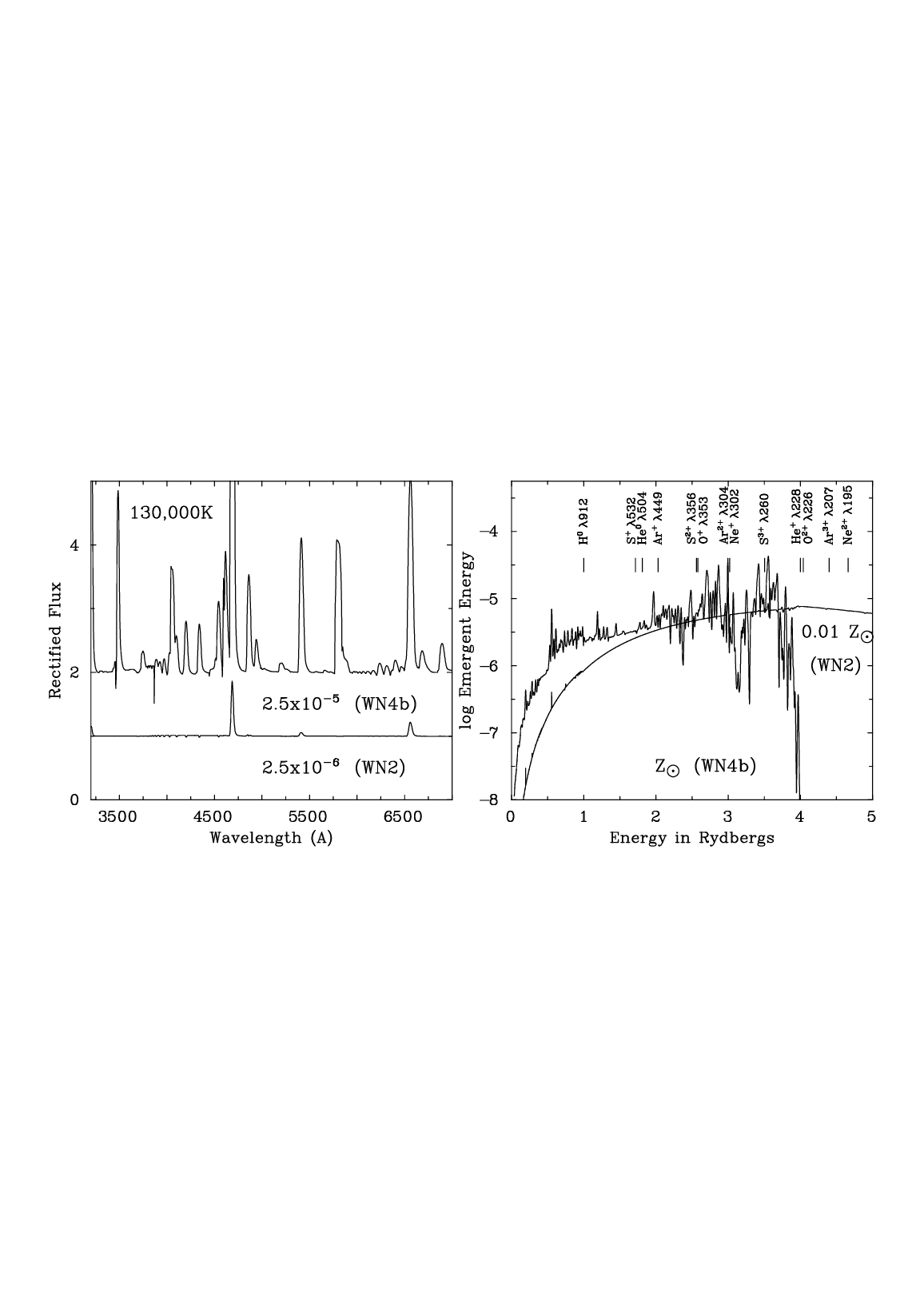}
\caption{Comparison between WNE models at fixed stellar parameters
(130kK), except that mass-loss rates (Fe-contents) differ by a 
factor of ten (100). The low mass-loss rate model
mimics a WN2-type spectral appearance, with a huge ionizing flux
in the He\,{\sc ii} continuum.}
\label{fig6}
\end{figure}

\subsection{Nebular He\,{\sc ii} $\lambda$4686 and bolometric corrections}

For my second case, I have taken the earlier 130kK WN model, with a
synthetic spectrum of a strong-lined WN4 star, and reduced its 
mass-loss rate by a factor of ten (scaling its metal content to 
0.01$Z_{\odot}$). The resulting optical spectrum
would be classified as a weak-lined WN2 star, as shown in Fig.~\ref{fig6}.
Its ionizing flux distribution is extremely hard, with a very strong flux 
above 54eV ($\sim$40\% of its entire luminosity!). {\it If mass-loss rates of 
WR stars are driven by 
radiation pressure, their spectral appearance and ionizing properties
will be very sensitive to metallicity.} The low metallicity WR model 
presented here may have application in very metal-poor starbursts, such as 
I~Zw~18 which is  thought to contain WR stars (de Mello  et al. 1998).

The above results suggest that solely hot WR stars with weak
winds produce a significant flux in the He$^{+}$ continuum, most likely
at low metallicities. This is supported by the known sample of WRs whose
nebulae show strong He\,{\sc ii} $\lambda$4686 emission by Garnett
et al. (1991), namely WR102 (WO, Galaxy), Brey~2 (WNE, LMC), Brey 40a (WNE+O,
LMC), AB7 (WNE+O, SMC), DR1 (WO, ICI613). Young, 
low metallicity starbursts would be expected to exhibit strong 
nebular He\,{\sc ii} $\lambda$4686 emission from WR stars, in contrast 
with high metallicity starbursts.

For the grid of high wind density models, 
representative of strong-lined Galactic WR stars, B.C's in the range 
$-$2.6 to $-$4.4 mag (WN), and $-$3.1 to $-$4.6 mag (WC) are obtained. 
Since wind density affects the ionizing spectrum of WR models, bolometric
corrections are also affected. B.C's for the WO and WN2 models are much higher
and very wind density sensitive, ($-$5.8 and $-$7.1 mag, respectively). 
Smith et al. (1994) used observations of clusters in the Galaxy to estimate WR 
masses and B.C's, namely $-$4.5 mag for WC stars, and $-$4 to $-$6 mag for WN 
stars, in fair accord with  predictions. Massey (these proc.) has repeated this for 
the LMC, and finds B.C's of $-6$ to $-$8 mag for cluster WNE stars. From calculations 
performed here, such stars would be expected to have low wind densities and emit 
strongly above 54eV. Detailed analysis of individual LMC WNE stars are 
sought in order to verify these predictions.

Overall, I have discussed the techniques used to derive stellar and
chemical properties of WR stars, highlighting the importance of clumping,
line blanketing on derived parameters, and the role of wind density and
metallicity on the emergent spectrum and ionizing properties of WR stars.

\acknowledgments
I would like to thank my collaborators, especially John Hillier.
Bill Vacca brought the importance of reliable ionizing fluxes of 
WR stars to my attention. PAC is a Royal Society University Research Fellow.

\end{document}